# New superconduting T'-La$^{3+}_{2-x}$RE$^{3+}_{x}$CuO$_4$ with isovalent doping ($RE$ = Sm, Eu, Tb, Lu, and Y)

Akio Tsukada, Yoshiharu Krockenberger, Hideki Yamamoto, and Michio Naito

NTT Basic Research Laboratories, NTT Corporation, 3-1 Wakamiya, Morinosato,

Atsugi-shi, Kanagawa 243-0198, Japan

We report the synthesis of new superconducting cuprates T'-La$_{2-x}$RE$_x$CuO$_4$ ($RE$ = Sm, Eu, Tb Lu, and Y) using molecular beam epitaxy. The new superconductors have no effective dopant, at least nominally. The substitution of isovalent $RE$ for La was essentially performed to stabilize the T' phase of La$_2$CuO$_4$ instead of the T phase. The maximum $T_c^{\text{onset}}$ is ~ 25 K and $T_c^{\text{zero}}$ is ~ 21 K. The keys to our discovery are (1) the preparation of high-crystalline-quality La-based T' films by low-temperature (~650°C) thin film processes, and (2) more thorough removal of impurity oxygen at the apical site, which is achieved by the larger in-plane lattice constant ($a_0$) of T'-La$_{2-x}$RE$_x$CuO$_4$ than other T'-$Ln_2$CuO$_4$ ($Ln$ = Pr, Nd, Sm, Eu, Gd) with the aid of large surface-to-volume ratio of thin films. The photoemission spectra suggest that the end member compounds of the T' system may not be a Mott insulator.

The lanthanide copper oxides of a general formula $Ln_2CuO_4$ ($Ln$ = lanthanide element) crystallize in two different structures: the $K_2NiF_4$ (T) structure with octahedral $CuO_6$ coordination and the $Nd_2CuO_4$ structure (T') with square-planar $CuO_4$. It has been believed that, in either case, the non-doped ($Cu^{2+}$) end member is a Mott insulator, and that superconductivity can be achieved only after $p$-type doping in the T structure or after $n$-type doping in the T' structure [1]. Against this belief, in this letter, we demonstrate that superconductivity is achieved in T'-$La_2CuO_4$ by substitution of $La^{3+}$ by "isovalent" rare earth ions ($RE^{3+}$) having a small ionic radius. This substitution is intended mainly to stabilize the T' phase of $La_2CuO_4$, so it is presumed not to change the Cu valence from 2+. Therefore, our observation appears to contradict the general belief that the end member of high-$T_c$ cuprates is a Mott insulator. Hence, it urges a reexamination of the early controversy on the "doped Mott-insulator scenario", which is now widely accepted as a starting viewpoint for high-$T_c$ superconductivity [2].

Superconductivity in T' cuprates is very sensitive to impurity oxygen at the O(3) site (apical site), whose presence is detrimental to achieving superconductivity. As-grown T' cuprates ($Ln_{2-x}Ce_xCuO_4$) are not superconducting [3,4] because they contain a fair amount of impurity oxygen atoms at the apical site. Superconductivity appears only after heat treatment to remove apical oxygen. In the "well-known" bulk phase diagram of T' superconductors, the superconducting (SC) region is adjacent to the antiferromagnetic (AF) region, and the superconductivity suddenly appears at the SC-AF boundary with maximum $T_c$ [5], suggesting competition between AF and SC orders. In the case of "optimally" doped $Pr_{1.85}Ce_{0.15}CuO_4$ or $Nd_{1.85}Ce_{0.15}CuO_4$, the AF correlations exist in as-grown non-superconducting samples, but they essentially disappear in reduced superconducting samples [6, 7]. This implies that more complete removal of apical oxygen would weaken the AF correlations and thereby expand the superconducting region. This was actually demonstrated by Brinkmann *et al.*, who achieved an extended superconducting region in $Pr_{2-x}Ce_xCuO_4$ (PCCO) using an improved reduction technique [8]. They wrapped the single crystals with two polycrystalline pellets from the top and bottom of the same composition and reduced them together at 1080°C for 3 days. After the reduction, they obtained superconducting PCCO with a wide Ce concentration range ($0.04 < x < 0.17$).

In previous articles [9, 10], we reported the thin-film synthesis of superconducting T'-$La_{2-x}Ce_xCuO_4$. This compound can be stabilized only at low synthesis temperatures of 600 ~ 700°C, so its bulk synthesis is difficult but its thin-film synthesis is rather easy. Compared with PCCO and $Nd_{2-x}Ce_xCuO_4$ (NCCO), the superconducting region of T'-$La_{2-x}Ce_xCuO_4$ extends to low doping, and the superconductivity appears at $0.045 < x < 0.22$. Furthermore, the end-member T'-$La_2CuO_4$ has low (2 mΩcm at 300 K) and metallic resisitivty down to 150 K, although superconductivity does not appear [11]. Our further efforts to improve metallicity toward a sign of superconducitivty in T'-$La_2CuO_4$ films were hampered by the upper limit of ~ 600°C in the synthesis temperature, which resulted in rather poor crystallinity. To see if we could stabilize the T' phase of $La_2CuO_4$ at > 600°C to improve crystallinity, we substituted large $La^{3+}$ by a small amount of small $RE^{3+}$. This led to the discovery of superconducting T'-$La_{2-x}RE_xCuO_4$ with $T_c$ ~ 20 - 25 K. The reason for the superconductivity seems to be that the large in-plane lattice constant ($a_0$) of T'-$La_{2-x}RE_xCuO_4$ enables more thorough removal of impurity oxygen at the apical site with the aid of large surface-to-volume ratio of thin films.

We grew La$_{2-x}$RE$_x$CuO$_4$ (RE = Pr, Nd, Sm, Eu, Tb, Lu, and Y) thin films in a customer-designed molecular beam epitaxy chamber from metal sources using multiple electron-gun evaporators [12]. The stoichiometry was adjusted by controlling the evaporation beam flux of each element using electron impact emission spectrometry via a feedback loop to the electron guns. During growth, 1 ~ 5 sccm of ozone gas (10% O$_3$ concentration) was supplied to the substrate for oxidation. The substrate temperature was typically ~ 650°C. The growth rate was ~ 1.5 Å/s, and the film thickness was typically ~ 900 Å. After the growth, the films were held at ~ 630°C in vacuum ($P_{O_2} < 10^{-8}$ Torr) for 10 minutes to remove interstitial apical oxygen. We mainly used SrTiO$_3$ (100) substrates but sometimes used KTaO$_3$ (100), NdCaAlO$_4$ (001), or YAlO$_3$ (100) substrates for the T' stabilization [11]. The lattice parameters and crystal structures of the grown films were determined using a standard X-ray diffractometer. Resistivity was measured by the standard four-probe method using electrodes formed by Ag evaporation.

First, we show the results for (La$^{3+}$,Y$^{3+}$)$_2$CuO$_4$, in which both La and Y undoubtedly have the valence of 3+. Figure 1(a) shows the $c$-axis lattice constant ($c_0$) of La$_{2-x}$Y$_x$CuO$_4$ films as a function of Y concentration $x$. At $x < 0.09$, the films stabilize in the T phase. At $x = 0.09$, the film is a two-phase mixture of T and T'. A further substitution of Y results in a single T' phase. In the T' phase, $c_0$ monotonically decreases with increasing $x$, which indicates that the substitution of La by Y in La$_{2-x}$Y$_x$CuO$_4$ is successful up to a solubility limit of $x = 1$. Figure 1(b) shows the temperature ($T$) dependence of resistivity ($\rho$) for the films of La$_{2-x}$Y$_x$CuO$_4$ with different $x$. The T-phase films of $x < 0.09$ are insulating. Metallic behavior (d$\rho$/d$T > 0$) is observed at $x = 0.09 - 0.5$.  Superconductivity appears at $x = 0.09$ and $0.15$.  Above $x = 0.15$, the resistivity increases with increasing $x$.

Similar investigations have been performed on other trivalent $RE$ ions (Pr, Nd, Sm, Eu, Tb [13], and Lu), and superconductivity was observed except for Pr and Nd substitution.  Figure 2 shows the $\rho$-$T$ curves for superconducting $La_{2-x}Sm_xCuO_4$, $La_{2-x}Eu_xCuO_4$, $La_{2-x}Tb_xCuO_4$, $La_{2-x}Lu_xCuO_4$, and $La_{2-x}Y_xCuO_4$ films.  The maximum $T_c$ for substitution by Sm, Eu, Tb, and Y is ~ 25 K at onset and ~ 20 K at zero resistance, which are close to the optimum $T_c$ of PCCO or NCCO.  The $T_c$ for Lu substitution is considerably lower, 18 K at onset and 12 K at zero resistance.  The superconductivity was also confirmed by substantial diamagnetic signal in magnetization measurements with the magnetic field parallel to the film surface, indicating the bulk nature of superconductivity in these compounds.

Figure 3 summarizes the $RE$ doping dependence of $T_c$ in T'-$La_{2-x}RE_xCuO_4$.  Here, $T_c$ is defined as the temperature of zero resistance.  One can see two trends in the figure: (1) the $T_c$ increases with decreasing $x$ until the T-to-T' boundary is encountered, and (2) the superconducting region becomes wider as the ionic radius of $RE$ increases from Lu to Sm.

The synthesis of T'- $La_{2-x}Y_xCuO_4$ [14, 15] by bulk processes has been reported by Takayama-Muromachi *et al.* and O *et al.*  Both employed a coprecipitation technique to obtain a finely mixed precursor, which is required for lowering the firing temperature to stabilize the T' phase.  In either attempt, no sign of superconductivity was observed.  The reasons for the contradicting observations may be, firstly, differences in the crystallinity and, secondly, differences in the concentration of residual apical oxygen atoms between our superconducting $La_{2-x}Y_xCuO_4$ films and their non-superconducting $La_{2-x}Y_xCuO_4$ bulk samples. Our superconducting films have better crystallinity and fewer apical oxygen atoms.

Figure 4 shows the *in-situ* valence band photoemission spectrum of superconducting $La_{1.85}Eu_{0.15}CuO_4$ ($T_c^{zero}$ = 20 K) in comparison with that of $La_{1.9}Ce_{0.1}CuO_4$ ($T_c^{zero}$ = 28 K) [16]. The measurements were performed at ambient temperatures with 50-eV photons. The films were transferred from the MBE chamber to the photoemission chamber in a portable vacuum container ($P \sim 10^{-8}$ Torr). Both spectra show a clear Fermi edge, and are essentially identical except for a rigid-band shift of ~ 0.2 eV. The experimental spectra are in good agreement with the density of states (DOS) calculated by the local density approximation (LDA) band calculations for $Nd_2CuO_4$ [17]. Furthermore, the observed rigid band shift of ~ 0.2 eV is also close to the value (~ 0.13 eV) predicted from the difference in the band filling [17] assuming that Eu valence is 3+ and Ce valence is 3.8+ to 4.0+.

Next, we discuss the origin of superconducting carriers in T'-$La_{2-x}RE_xCuO_4$. We can think of two possible scenarios. One is that oxygen deficiencies at the O(2) site are a source of effective electron carriers. With regard to this possibility, neutron diffraction experiments provide information about the site-specific occupancy of oxygen. All the neutron diffraction experiments performed on T'-$Nd_2CuO_4$ or T'-$(Nd,Ce)_2CuO_4$ qualitatively agree with one another in that the reduction procedure mainly reduces the impurity O(3) occupancy but does not change the O(1) and O(2) occupancies within $\delta = \pm 0.01$ [18-20]. Quantitatively, however, the reported occupancy varies report by report, from 1.94 to 2.00 for O(1) and from 1.90 to 2.00 for O(2). The presence of O(1) deficiencies in our superconducting films is unlikely since it is believed that the $CuO_2$ planes should be perfect for achieving high-$T_c$ superconductivity. The possibility of O(2) deficiencies is also not unambiguously supported from the neutron diffraction experiments, but cannot be excluded. This has to be examined further, although there is no established technique for determining the site-specific occupancy of oxygen in thin films.

The other is that T'-$La_2CuO_4$ is not a Mott insulator, and has intrinsic carriers. Band calculations using the LDA [linear augmented plane-wave (LAPW) or Linear muffin-tin-orbital (LMTO)] method predict that both T-$La_2CuO_4$ and T'-$Nd_2CuO_4$ have a metallic band structure [17, 21, 22]. The apparent failure of the band calculations in predicting the antiferromagnetic insulating ground state in T-$La_2CuO_4$ has led to a universal view that the normal state of cuprate superconductors may be a non-Fermi liquid described as a doped Mott-insulator rather than a normal band metal. In our MBE-grown films, however, the physical properties of T-$La_2CuO_4$ and T'-$La_2CuO_4$ show a striking contrast [11]. With no excess oxygen provided, T-$La_2CuO_4$ is insulating with $\rho$(300 K) ~ 100 $\Omega$cm, whereas T'-$La_2CuO_4$ is metallic at room temperatures with $\rho$(300 K) ~ 0.002 $\Omega$cm and weakly localized at low temperatures. Additionally, our *in-situ* ultraviolet photoemission spectra showed no Fermi edge in T-$La_2CuO_4$, whereas a clear Fermi edge was observed in T'-$La_2CuO_4$ [23]. Both of these results indicate that T-$La_2CuO_4$ has an insulating ground state and T'-$La_2CuO_4$ has a metallic ground state, and hence are examples against the universal view that the end member of high-$T_c$ cuprates is a Mott insulator. The dramatic difference between T-$La_2CuO_4$ and T'-$La_2CuO_4$ seems to arise from the presence or absence of the apical oxygen. Some band calculations have predicted a significant change in the electronic structure due to the presence of apical oxygen [22, 24]. That the main features of the experimental photoemission spectra, including the rigid band shift between T'-(La,Eu)$_2$CuO$_4$ and T'-(La,Ce)$_2$CuO$_4$ (Fig. 4), can be reproduced by the band calculations also strongly supports the second scenario.

The "apparent" insulating behavior in the end-member T' compounds is not intrinsic but arises from the localization of carriers due to residual apical oxygen atoms. Therefore, the $\rho$-$T$ behavior depends strongly on the concentration of residual apical oxygen atoms. With the same compound, the $\rho$-$T$ becomes more metallic with more complete removal of apical oxygen [25]. With the same reduction recipe, the resistivity of T'-$Ln_2$CuO$_4$ becomes more metallic with larger $Ln$ ions. The latter trend is demonstrated in Fig. 5, and can be understood on the basis of the T' family's solid state chemistry, which indicates that the concentration of residual oxygen atoms, with the same reduction recipe, decreases with an increasing $Ln$ ionic radius. Figure 5 plots $\rho$(50 K) of T'-La$_{2-x}$RE$_x$CuO$_4$ as a function of the effective ionic-radius ($r_i^{\text{eff}}$) of La$_{2-x}$RE$_x$ and includes the data for pure T'-$Ln_2$CuO$_4$ films ($Ln$ = La, Pr, Nd, Sm Eu, and Gd). The $\rho$(50 K) is lower for a larger ionic radius. Then, the condition necessary for the superconductivity to be achieved is $\rho$(50 K) < ~ 0.5 - 1 mΩcm. Finally, we should note that the superconductivity was not achieved in T'-(La,Pr)$_2$CuO$_4$ or T'-(La,Nd)$_2$CuO$_4$, although their resistivity is low enough to satisfy the above criterion. That we have not yet observed superconductivity in these compounds could be due to our insufficient efforts, but we have noticed that the metallicity in these two compounds is weaker than that in other superconducting compounds, although the absolute resistivity value is low.

In summary, we prepared new superconducting T'-La$_{2-x}$RE$_x$CuO$_4$ ($RE$ = Sm, Eu, Tb, Lu, and Y) thin films with no effective dopant, at least nominally. The maximum $T_c^{\text{onset}}$ of ~ 25 K and $T_c^{\text{zero}}$ of ~ 21 K are obtained at the T-T' phase boundary.

The larger $a_0$ of T'-$La_{2-x}RE_x$CuO$_4$ than that of other T'-$Ln_2$CuO$_4$ enables more thorough removal of impurity oxygen at the apical site, which leads to the superconductivity. The *in-situ* valence band photoemission results imply that the electronic structure of these compounds is well described within the framework of band theory. Our results suggest that the end member compounds of T'-$Ln_2$CuO$_4$ are not Mott insulators.

**Acknowledgments**

The authors thank Dr. T. Yamada, Dr. A. Matsuda, Dr. H. Sato, Dr. S. Karimoto, Dr. K. Ueda, and Dr. J. Kurian for helpful discussions, and Dr. M. Morita, Dr. H. Takayanagi, and Dr. S. Ishihara for their support and encouragement throughout the course of this study.

[13] Here we discuss the Tb valence. Tb takes trivalent and also tetravalent states. Then Tb valence was analyzed by *c*-axis variation and XPS measurement. The *c*-axis lattice constant of $La_{2-x}Tb_xCuO_4$ monotonically decreased with increasing $x$ and

smoothly connected to the reported value of bulk $Tb_2CuO_4$ at $x = 2$ [P. Bordet *et al.*, Physica C **193**, 178 (1992)]. This result indicates that Tb ion size is close to trivalent ion size (1.04 Å) rather than tetravalent ion size (0.88 Å). Comparison of XPS spectra at Tb-4d states between our film at $x = 0.3$ and references ($TbO_2$ and $Tb_2O_3$) [D. D. Sarma *et al.*, J. Electron Spectrosc. Relat. Phenom. **20**, 25 (1980)] also indicate Tb valence is 3+ or close to 3+.

**Figure captions**

Figure 1  (a) $c$-axis lattice constant ($c_0$) of $La_{2-x}Y_xCuO_4$ films as a function of Y concentration.  Square and circles represent T and T'-phases, respectively.  The gray area at $x < 0.09$ indicates the region where the T phase stable.  The symbols connected by vertical dotted line indicate multiple-phase formation.  Solubility limit of Y in $La_{2-x}Y_xCuO_4$ is $x = 1$.  (b) Temperature dependence of resistivity for $La_{2-x}Y_xCuO_4$ films with different $x$.  Superconductivity appears at $x = 0.09$ and 0.15.

Figure 2  The best $\rho$-T curves for films of $La_{2-x}RE_xCuO_4$ ($RE$ = Sm, Eu, Tb, Lu and Y).  Circles, squares, triangles, crosses, and diamonds are for films with Sm, Eu, Tb, Lu, and Y, respectively.  The inset shows superconducting transitions.

Figure 3  The $RE$ concentration dependence of $T_c$ for $La_{2-x}RE_xCuO_4$ films.  $T_c$ is defined as the temperature of zero resistance (closed circles: Sm, closed squares: Eu, closed triangles: Tb, open squares: Lu, open circles: Y).

Figure 4  *In-situ* photoemission spectra of (a) $La_{1.85}Eu_{0.15}CuO_4$ and (b) $La_{1.9}Ce_{0.1}CuO_4$ thin films.  The spectra are integrated approximately in the shaded area in the momentum space.  The vertical axis is normalized by the intensity of the peak (BE ~3.5 eV).  Inset displays the near-$E_F$ region.

Figure 5  Resistivity at 50 K [$\rho$(50 K)] versus effective ionic-radius $r_i^{eff}$ defined as $[r_i(La^{3+})*(2-x) + r_i(RE^{3+})*x]/2$.  Each symbol assigned to Pr, Nd, Sm, Eu, Tb, Lu, and Y denotes the data of corresponding T'-$La_{2-x}RE_xCuO_4$ thin films ($RE$ = Pr, Nd, Sm,

Eu, Tb, Lu, and Y).    Crosses for EM represent data for the end-member T'-$Ln_2CuO_4$ thin films ($Ln$ = La, Pr, Nd, Sm, Eu, and Gd).    Open and closed symbols indicate non-superconducting and superconducting samples, respectively.

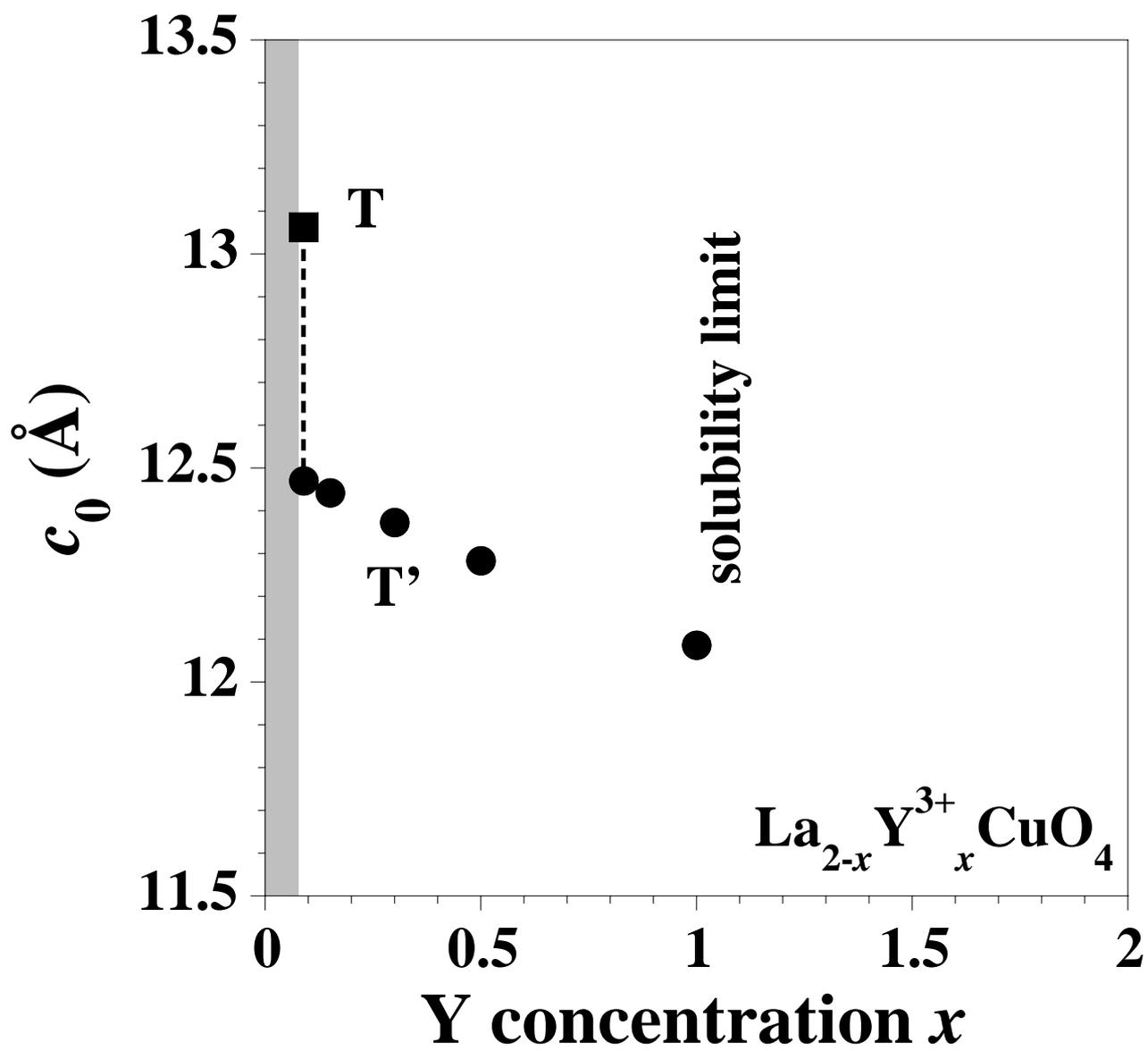

Figure 1(a). A. Tsukada *et al.*

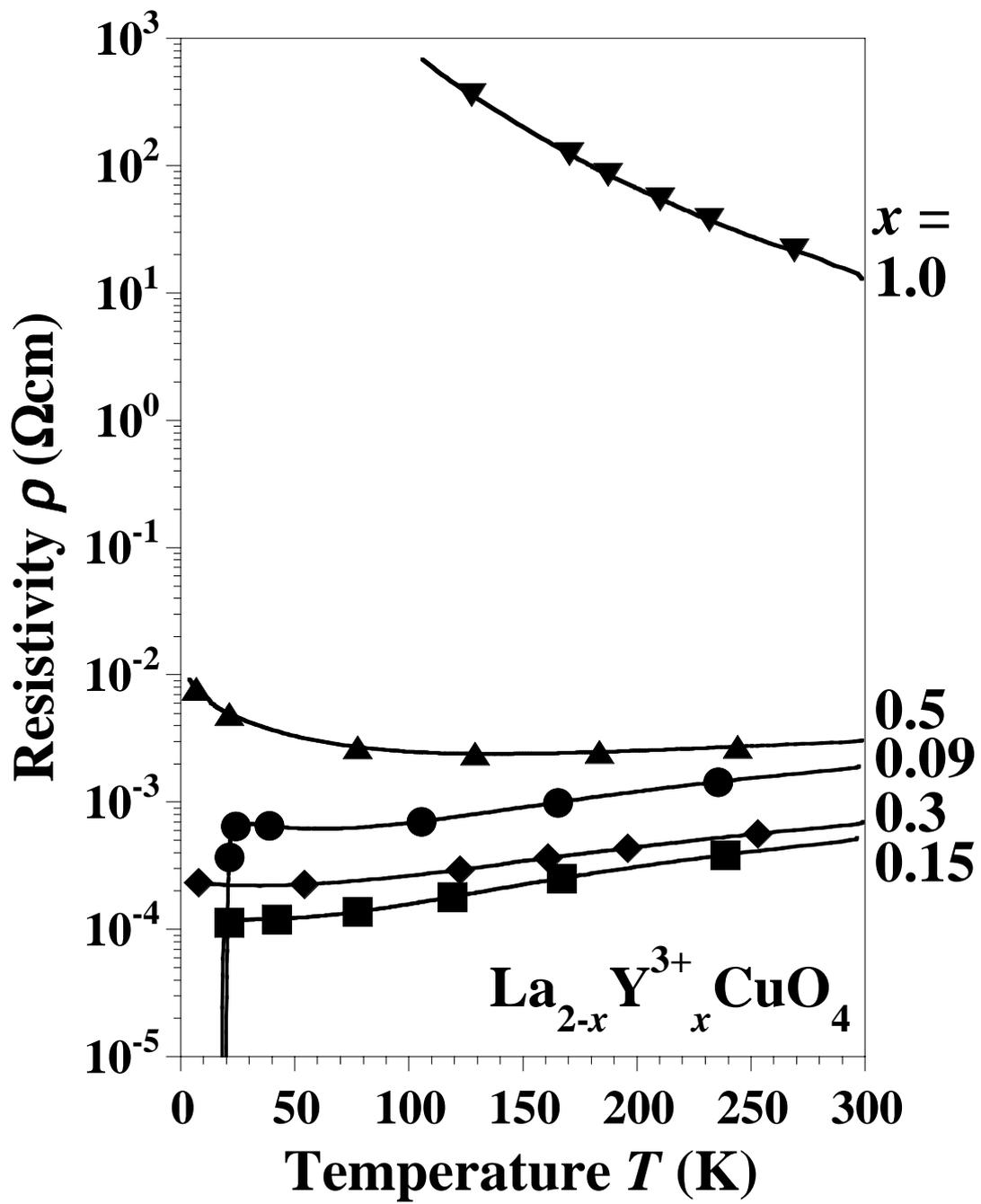

Figure 1(b). A. Tsukada *et al.*

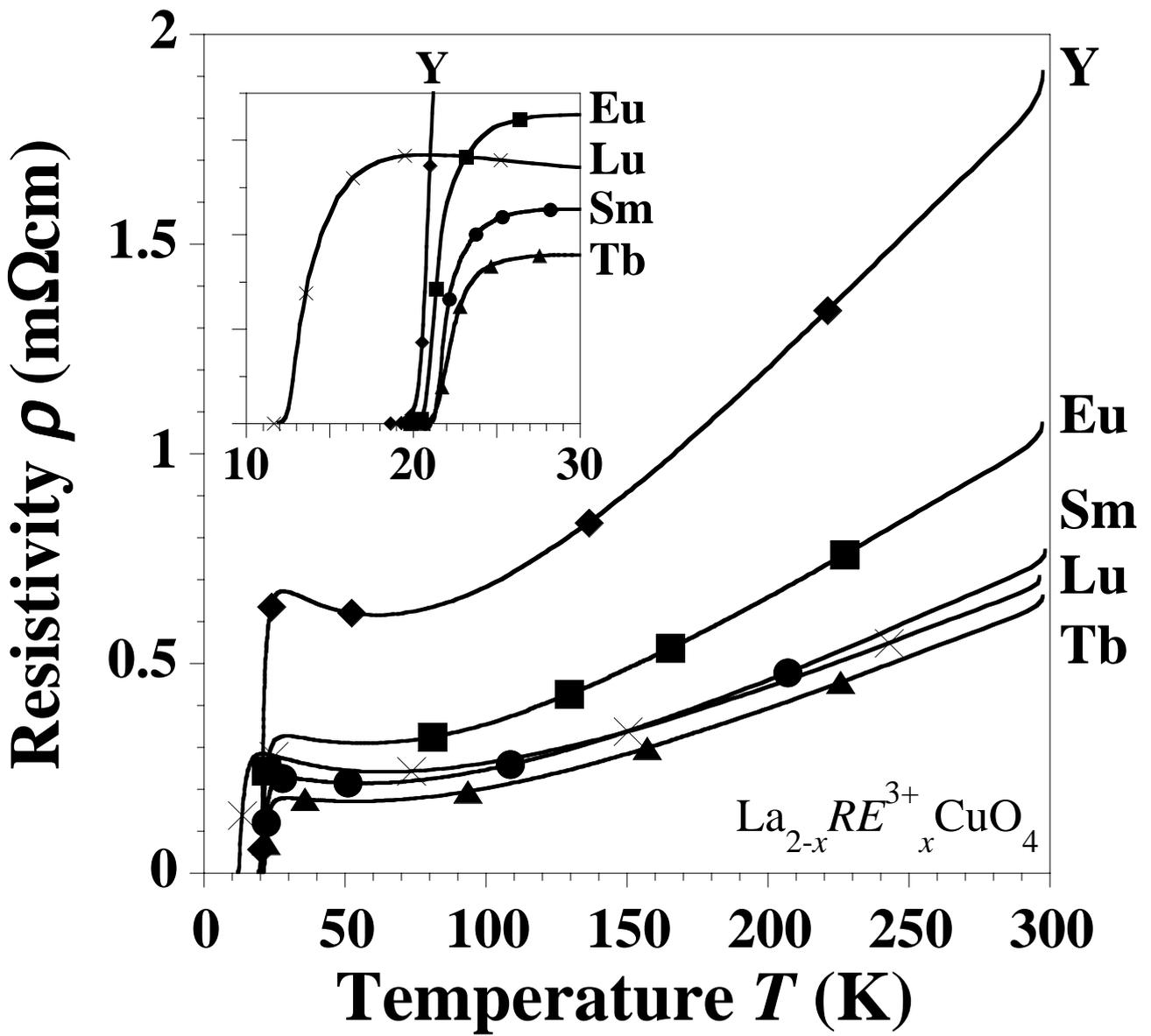

Figure 2. A. Tsukada et al.

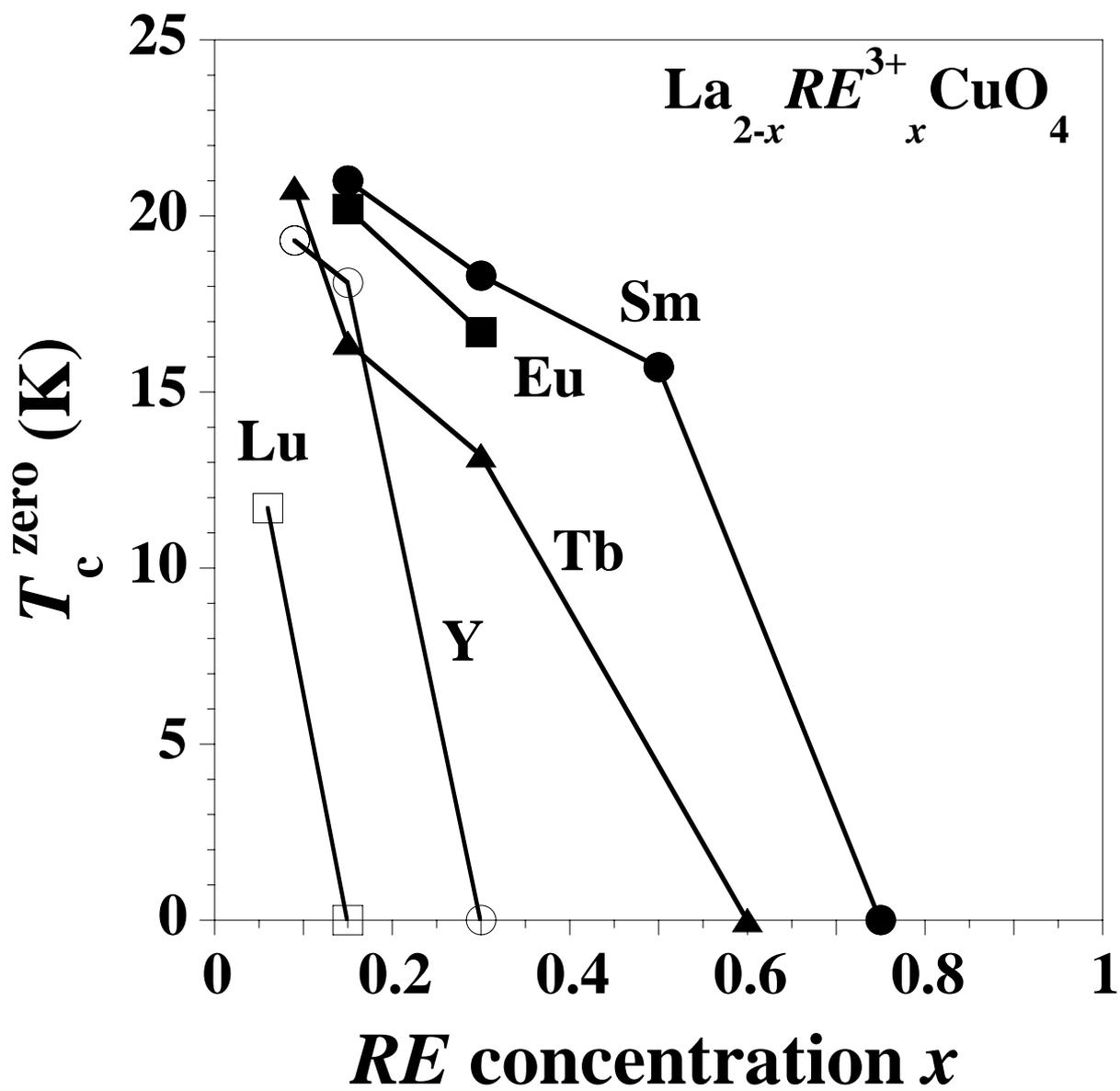

Figure 3. A. Tsukada *et al.*

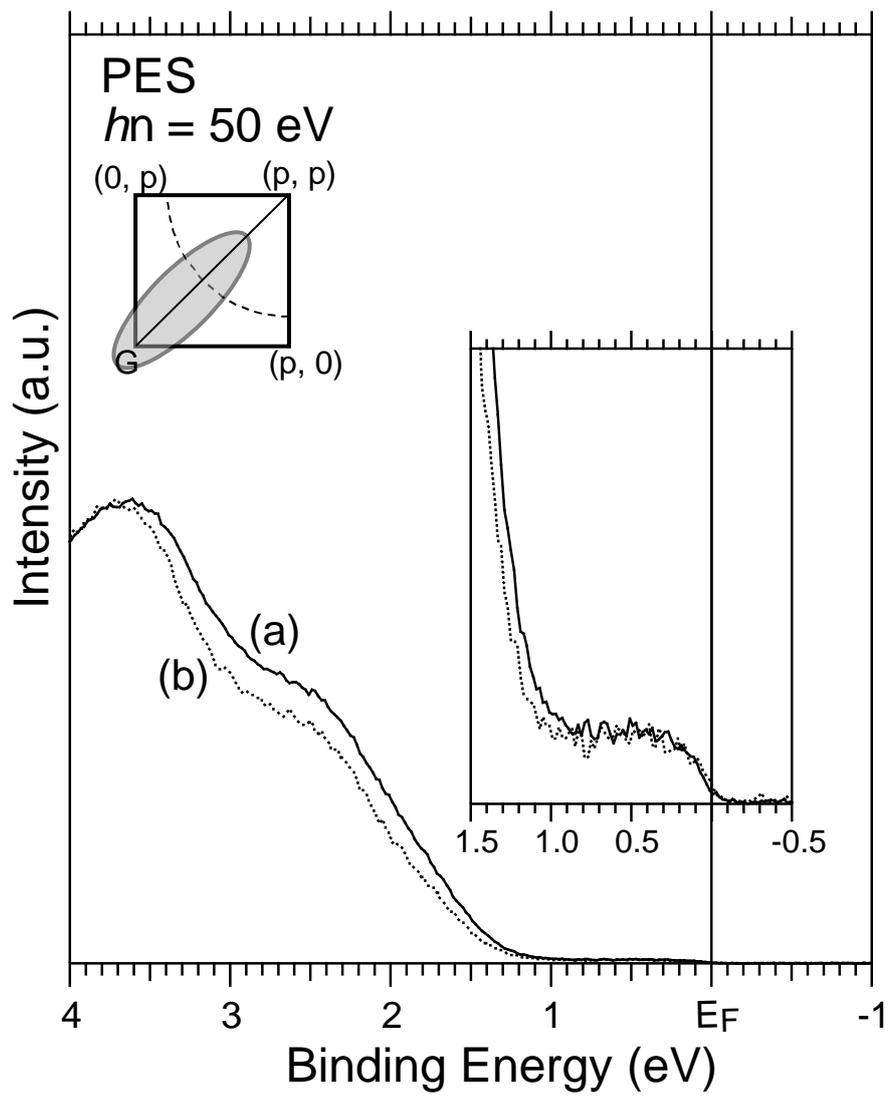

Figure 4. A. Tsukada *et al*.

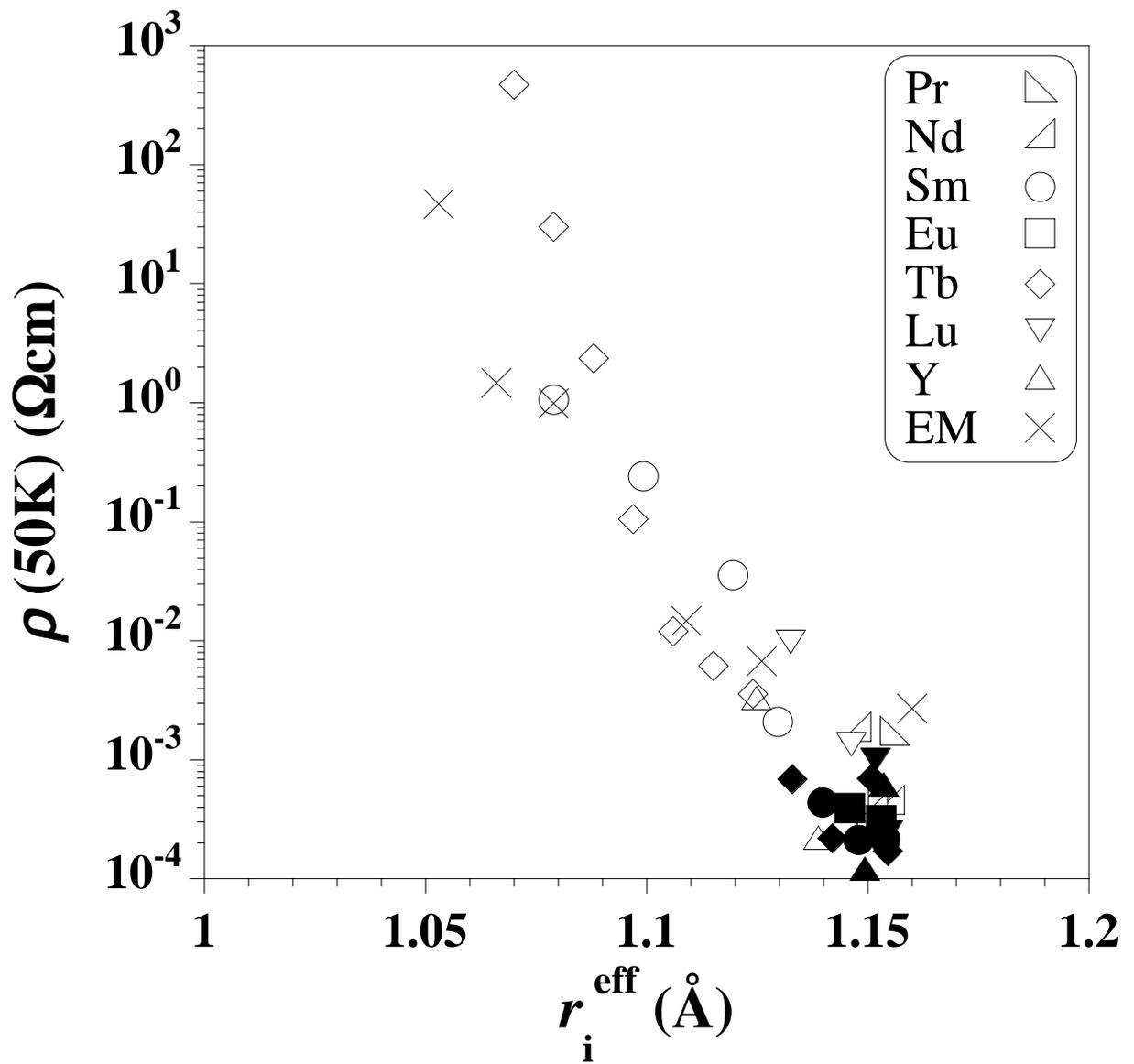

Figure 5. A. Tsukada *et al.*